\documentclass[twoside]{article}

\usepackage[sc]{mathpazo} 
\usepackage[T1]{fontenc} 
\usepackage{microtype} 
\usepackage{url}
\usepackage[applemac]{inputenc}
\usepackage[hmarginratio=1:1,top=32mm,columnsep=20pt]{geometry} 
\usepackage{multicol} 
\usepackage{graphicx}
\usepackage{epstopdf}
\usepackage{epsfig}
\usepackage{gensymb}
\usepackage[hang, small,labelfont=bf,up,textfont=it,up]{caption} 
\usepackage{booktabs} 
\usepackage{float} 
\usepackage{natbib}
\usepackage{lettrine} 
\usepackage{paralist} 

\usepackage{abstract} 
\usepackage{titlesec} 
\renewcommand\thesection{\Roman{section}}
\titleformat{\section}[block]{\large\scshape\centering}{\thesection.}{1em}{} 

\usepackage{fancyhdr} 
\title{\vspace{-15mm}\fontsize{24pt}{10pt}\selectfont\textbf{First deep underground observation of rotational signals from an earthquake at teleseismic distance using a large ring laser gyroscope}} 

\date{}

\begin{document}

\maketitle 

\thispagestyle{fancy} 
\vspace{-5em}
\centerline{\textbf{\underline{Andreino Simonelli}$^{1,2,3,\ast}$, {Jacopo Belfi$^1$}, {Nicolò Beverini$^{1,2}$}, {Giorgio Carelli$^{1,2}$}, {Angela Di Virgilio$^1$}} }\centerline{\textbf{{Enrico Maccioni$^{1,2}$}, {Gaetano De Luca$^{5}$}, {Gilberto Saccorotti$^{{4}}$}}}

\centerline{$^{\ast}$\emph{andreino.simonelli@pi.infn.it}}

\vspace{0.5cm}
\centerline{$^1$\emph{INFN section of Pisa, Largo B. Pontecorvo 3 56127, Pisa, Italy}}
\centerline{$^2$\emph{University of Pisa. Dept. of Physics ``E.Fermi'', Largo B. Pontecorvo 3 56127, Pisa, Italy}}
\centerline{$^3$\emph{Ludwig-Maximilians-University Munich, Theresienstra{\ss}e 41, D-80333 Munich, Germany}}
\centerline{$^4$\emph{INGV section of Pisa, Via della Faggiola, 32 - 56126 Pisa, Italy}}
\centerline{$^5$\emph{INGV-CNT Roma, Italy}}


\begin{abstract}

\noindent
Recent advances in large ring laser gyroscopes (RLG) technologies opened the possibility to observe rotations of the ground with sensitivities up to $10 ^{-11}$ rad/sec over the frequency band of seismological interest (0.01-1Hz), thus opening the way to a new geophysical discipline, i.e.  rotational seismology. A measure of rotations in seismology is of fundamental interest for (a) the determination of all the six degrees of freedom that characterize a rigid body's motion, and (b)  the quantitative estimate of the rotational motions contaminating ground translation measurements obtained from standard seismometers. Within this framework, this paper presents and describes  GINGERino, a new large observatory-class RLG located in Gran Sasso underground laboratory (LNGS), one national laboratories of the INFN (Istituto Nazionale di Fisica Nucleare). We also report unprecedented observations and analyses of the roto-translational signals from a tele-seismic event observed in such a deep underground environment.

\end{abstract}


\begin{multicols}{2} 

\section{Introduction}
\lettrine[nindent=0em,lines=3]
Ring Laser Gyroscopes (RLG) are the best sensors for capturing the rotational motions associated with the transit of seismic waves, thanks to the optical measurement principle, these instruments are in fact insensitive to translations. The claim for a rotational sensor in geophysics is outlined in a fundamental text about seismology \cite{aki}, where the authors state that ''... note the utility of measuring rotation near a rupturing fault plane (...), but as of this writing, seismology still awaits a suitable instrument for making such measurements ''. The search for such a sensor is of actual interest, as shown by many recent studies \cite{kalab2013application,brokevsova2010new,schreiber2006ring}. Nowadays RLGs allowed to achieve important results, spanning from geodesy \cite{schreiber2004direct}  to the analysis of earthquakes recorded over a wide range of distances \cite{igel2005rotational,pancha2000ring,miatesi,schreiber2006ring}.
The size or RLG changes, depending on the scope, from some centimeters to more than four meters.
RLGs for navigation are very small and lightweight; they are produced commercially and are widely adopted for either underwater or airborne platforms. Their sensitivity, however, is not sufficient for geophysical applications. Sensitivity and accuracy of RLGs increase with size, thus maximizing dimensions causes a minimization of physical effects that cause the gyro to work out of an ideal linear regime. Scientific results like the solid tides monitoring or a measure of the length of the day (LOD) are only achievable by very large frame RLG. Actually, the G-ring apparatus in Wettzel Germany represents the reference RLG for geodetic and seismological observations. Smaller in size and less expensive is the range of RLG of the class Geosensor, \cite{schreiber2006geosensor,belfi20121,belfi2012performance}. The GINGERino apparatus funded by INFN in the context of a larger project of fundamental physics is intended as a pathfinder instrument to reach the high sensitivity needed to observe general relativity effects; more detail are found at the URL (\url{https://web2.infn.it/GINGER/index.php/it/} and in \cite{2016arXiv160102874B}. 
\section{Instrumental apparatus}
\begin{figure}[H]
\begin{center}
\includegraphics[scale=.15,angle=0]{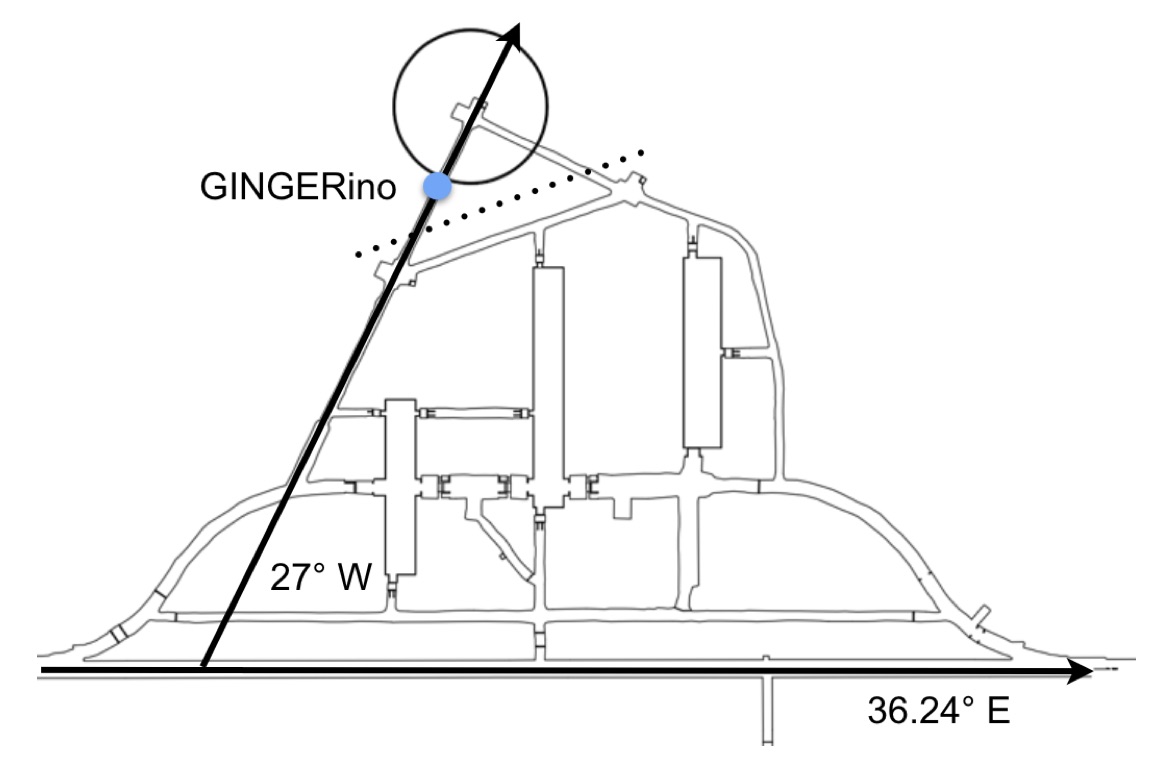}
\caption{Map of the LNGS underground laboratories}
\label{map}
\end{center}
\end{figure}
\begin{figure}[H]
\begin{center}t
\includegraphics[scale=.12,angle=0]{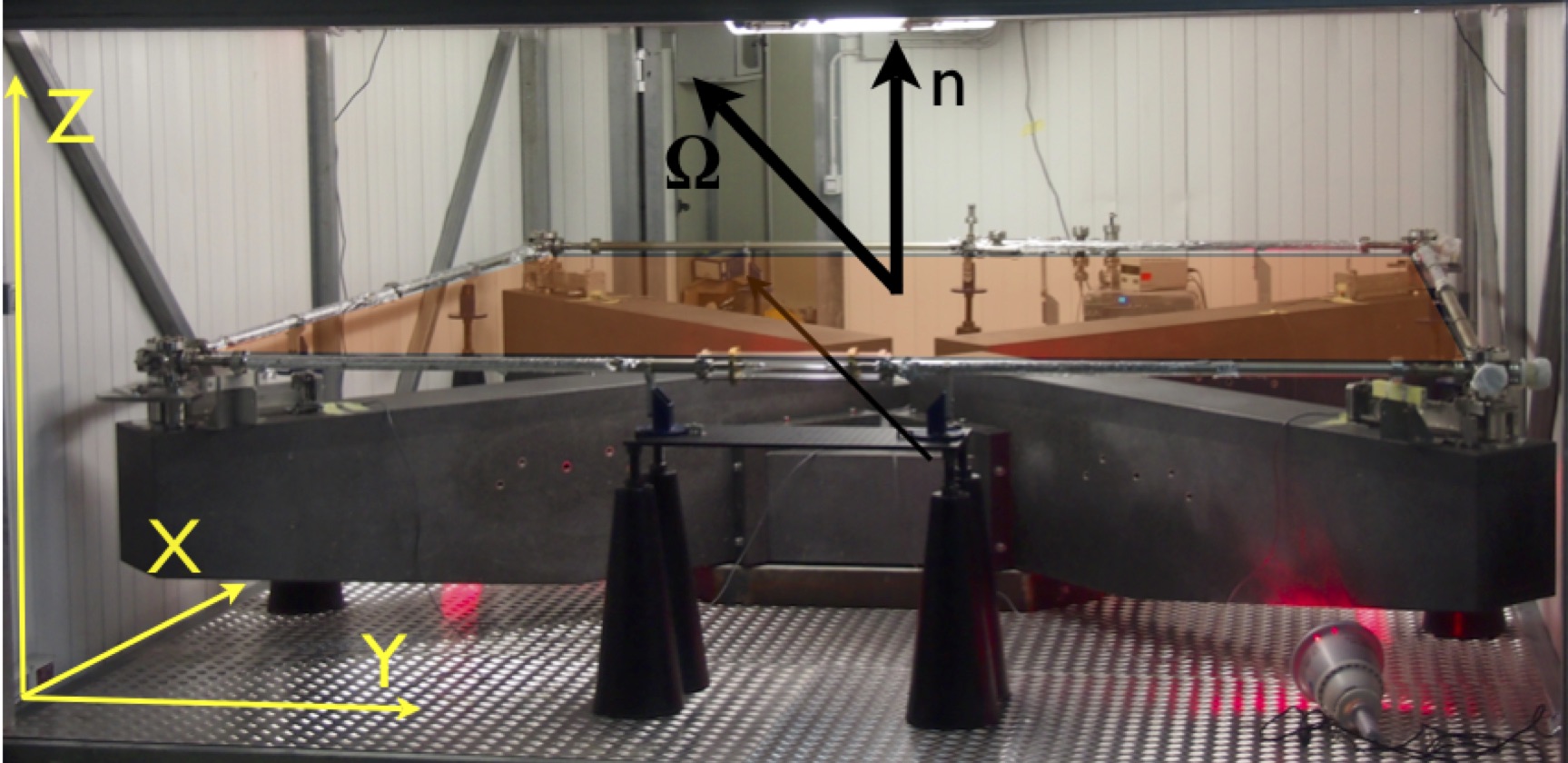}
\caption{The GINGERino RLG}
\label{gingerino}
\end{center}
\end{figure}
\begin{figure}[H]
\begin{center}
\includegraphics[scale=.13,angle=0]{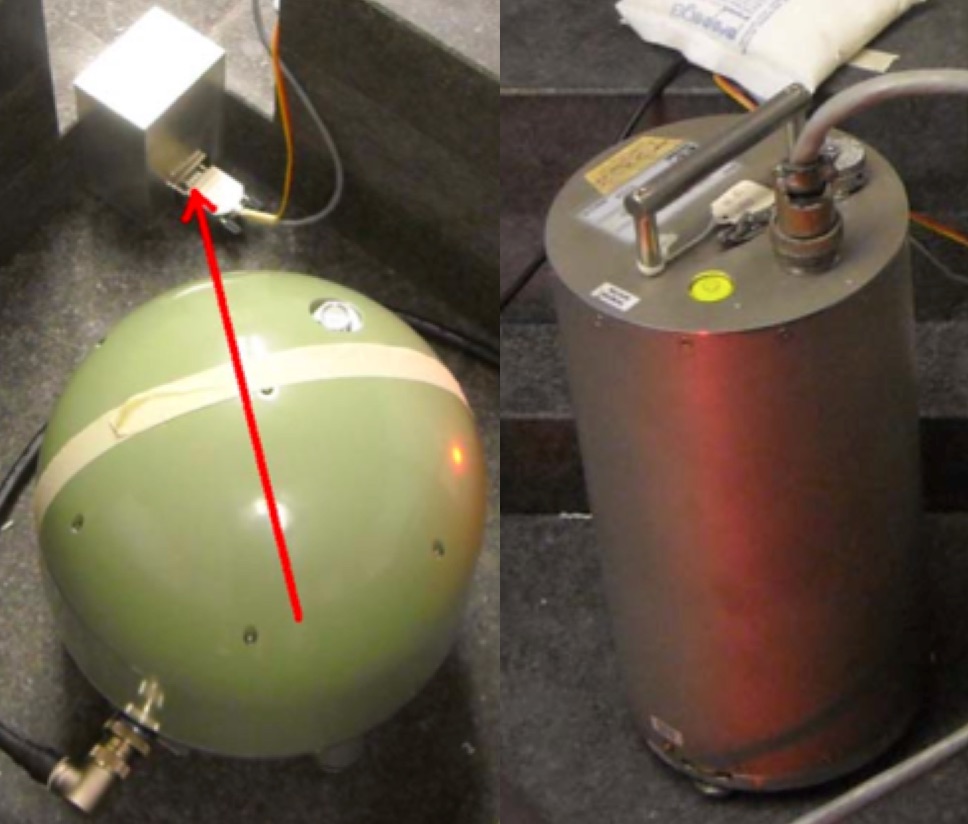}
\caption{The NANOMETRICS Trillium 240 s (left) and Guralp CMG 3T360s (right) and the Lippmann 2-K tilt meter (on top), the red arrow shows the North direction}
\label{trillamelo}
\end{center}
\end{figure}

The GINGERino is located Inside the Gran Sasso National Laboratory (LNGS) of the INFN (Fig. \ref{map}). The equipment of geophysical and seismological interest is constituted by the following instruments: The large He:Ne ring laser visible in Fig. \ref{gingerino}; this is a 3.6 m side square cavity ring laser installed over a granite structure block anchored to the rock of the B knot tunnel of the LNGS. This is our rotation sensor, it is able to detect rotations around the symmetry axis (oriented vertically) with a sensitivity better than $10^{-10}$ rad/s in the band of interest for global seismology (5 Hz-300s). A Nanometrics Trillium 240s seismometer which is installed at the center of the RLG granite frame Fig. \ref{trillamelo}. 
This instrument is part of the national earthquake monitoring program of the Istituto Nazionale di Geofisica e Vulcanologia (INGV hereinafter), provides the ground translation data to be compared to the RLG rotational data in order to infer the phase velocity measurements during the transit of shear and surface waves from earthquakes at local, regional and tele seismic distances. Further details on this station are at the URL (\url{http://iside.rm.ingv.it/iside/standard/info_stazione.jsp?page=sta&sta=2571}). A Lippmann 2-K digital tilt-meter with a resolution better then one nrad is placed beside the seismometer in order to monitor the possible slow ground tilt related to either local or wide scale (solid earth tides) effects. A second broadband seismometer, Guralp CMG 3T–360s (Fig. \ref{trillamelo}) is placed in the central block for data redundancy.

\section{Method}
RLG are based on the Sagnac effect; this effect is caused by a difference in the optical path as seen by two counter propagating laser beams that leads to a difference in the optical frequency between the clockwise and anti-clockwise propagating beams. The two beams are mixed out of the optical cavity in order to reveal the beat of the two slightly different frequencies. The beat frequency $f$, also called the Sagnac frequency can be related to the rotation rate around the normal vector to the surface outlined by the square optical path (see Fig. \ref{gingerino})
using the simple following equation: 
\begin{equation}
\Omega=\frac{\lambda_{He:Ne}}{L\sin\theta}f
\label{scalef}
\end{equation} where $\lambda_{He:Ne}$ is the wavelength of the He:Ne laser (632 nm), $L$ is the square side length and $\theta$ is the angle between the versor $\hat{n}$ and $\vec\Omega$. We know from theory \cite{aki} that rotations can be retrieved from ground displacement as the curl of the wave-field.
\begin{equation}
\vec\Omega=\frac{1}{2}(\nabla \times \vec u)
\label{curl}
\end{equation}
Referring to our setup (Fig. \ref{gingerino}) for example, the displacement caused by a Love wave traveling as a plane wave along the $x$-direction is expressed through the equation:
\begin{equation}
u_{y}=Ae^{i\omega(x/C_{L}-t)}
\label{wave}
\end{equation}
By applying eq. \ref{wave} to eq.\ref{curl} we obtain the relationship:
 \begin{equation}
\Omega_{z}=\frac{\ddot u_{y}}{2c_{L}}
\label{velfas}
\end{equation}
which provides a direct estimation of the phase velocity $C_{L}$ by using only a single-site measurement. From this latter formulation it is also evident that the sensing of ground rotations over the seismic frequency band requires high sensitivity: the phase-velocity scaling implies in fact that ground rotations are two to three orders of magnitude smaller than the associated translational movements. For this purpose a very sensitive and completely decoupled from translations device is required and at present large RLGs are the best candidates.
\section{First results}
\begin{figure}[!h]
\includegraphics[width=\textwidth]{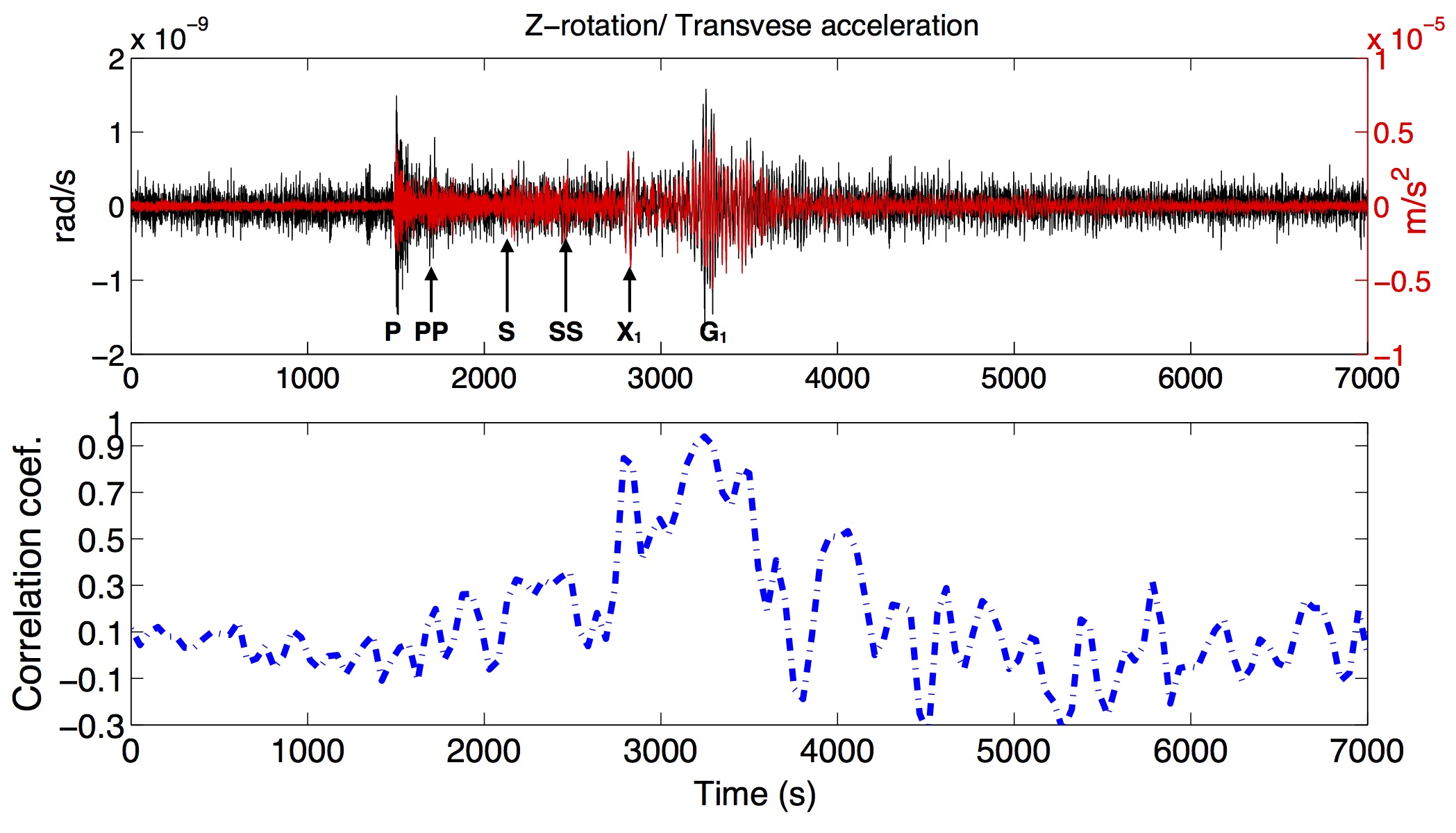}
\captionof{figure}{(top) Ground rotation and transverse acceleration time histories (black and blue lines, respectively), time zero is at 12:40:00 UTC. (bottom) ZLCC between the above traces. }
\label{rotocorr}
\end{figure}

\begin{figure}[!h]
\centering
\includegraphics[width=\textwidth]{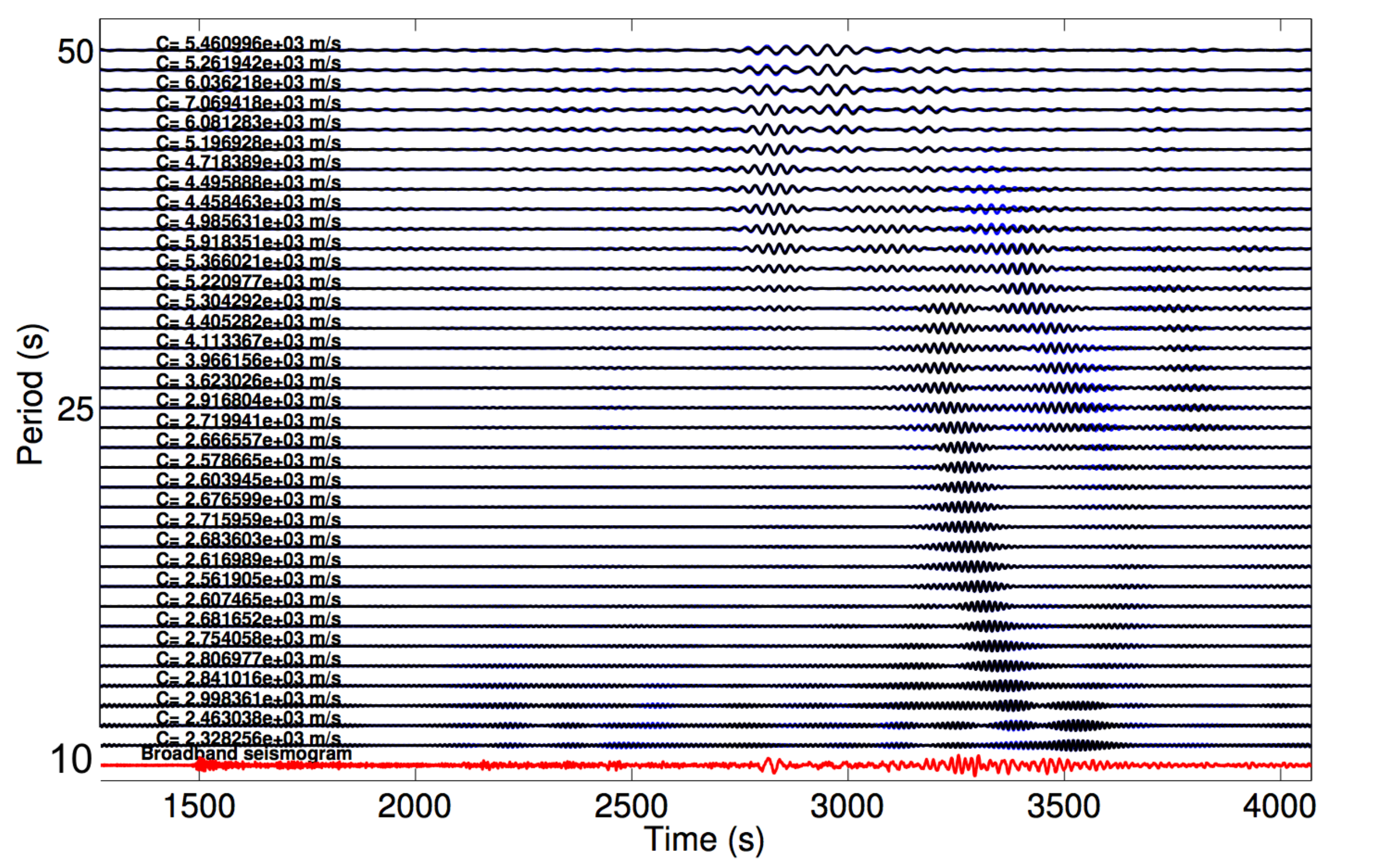}
\captionof{figure}{Superposition of trace-by trace normalized narrow band filtered signals, for every dominant period we report the estimated phase velocity}
\label{domram}
\end{figure}

An earthquake with magnitude 7 occurred on 17-06-2015  12:51:32 (UTC) with epicenter in the Southern Mid Atlantic Ridge [Sea] has been recorded by our instruments during the longest run of continuous data acquisition from 11/6/15 to 19/6/15. Though the recordings exhibit a poor signal-to-noise-ratio (SNR) their quality is sufficient to perform some analysis of seismological interest. The processing steps have been:\begin{itemize}
\item The N-S and E-W seismometer traces are rotated by a step of $1\degree$ over the $\{0, 2\pi\}$ range and for each rotation step, the zero-lag-cross-correlation (ZLCC) between the rotational signal and transverse accelerations is calculated. The maximum is found at a rotation angle of $198\degree$ N, the theoretical azimuth derived from epicenter and station coordinates is $202\degree$ N. The discrepancy between the observed and theoretical azimuthal values is small, once considering possible seismometer misorientation and deviation of surface wave trajectories from the great circle path as a consequence of lateral velocity heterogeneities.
\item The ZLCC between translational and rotational traces is calculated using a 200-seconds-long window, sliding with 50\% overlap. The Love-wave arrival is marked by a clear correlation peak (see Fig.\ref{rotocorr})
\item Ground rotations and Transverse accelerations (respectively blue and black lines in Fig. \ref{domram}) are narrow band filtered with a FIR filter with a 1 s large passband region form 1 s to 50 s of Period. In the frequency bands where ZLCC is above a threshold of 0.7, the amplitude ratio between the maxima of the envelopes evaluated via Hilbert transform gives a direct measure of phase velocity for that particular period (see Fig. \ref{domram}).
\end{itemize}
\end{multicols}
\begin{figure}[!h]
\centering
\includegraphics[scale=.19]{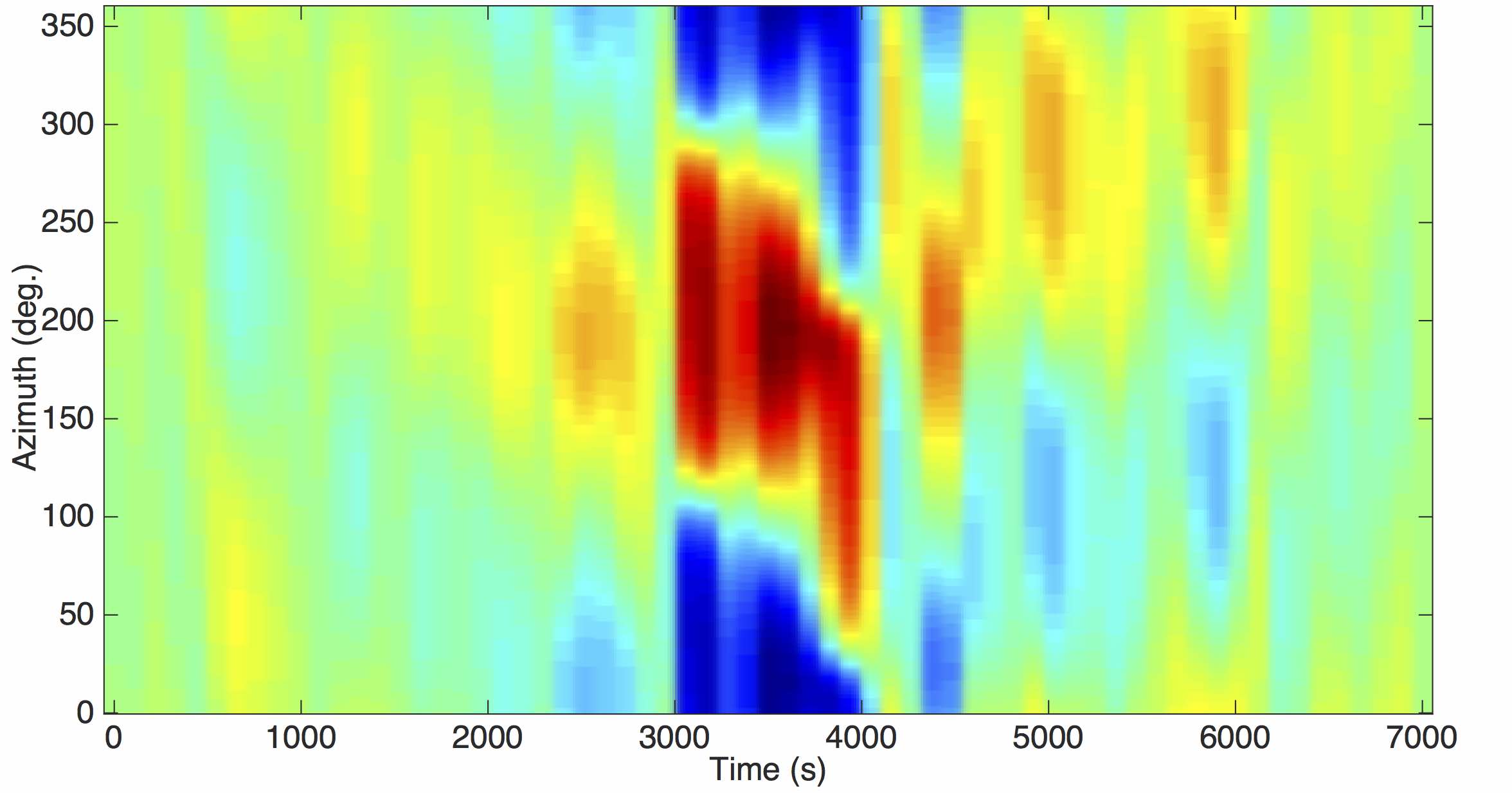}
\caption{Map of correlation versus rotation of transverse acceleration}
\label{sgoraha}
\end{figure}
\begin{multicols}{2}
\section{Conclusions}
GINGERino is a test apparatus, and improvements in sensitivity and stability of the apparatus are foreseen in the near future. At present the RLG is running in a preliminary test mode in order to optimize the experimental parameters that will allow us to let it run continuously together with tilt-meters and seismometers and to increase sensitivity in order to be able to detect the secondary microseism peak that is only a factor five below our noise floor at the 10 seconds period. In a previous study we used a smaller RLG oriented along an horizontal axis and we obtained consistent estimates of ground rotations associated with the transit of Rayleigh waves from the 2011, Mw=9.0 Japan earthquake \cite{GpisaJap}. The present availability of a larger and much more sensitive RLG as GINGERino now permits extending the analysis to earthquake signals over a wider magnitude range. The simultaneous measurement of ground translation and rotation of these sources will allow the definition of the dispersion curve of Love waves over a broad frequency range, from which a local shear-wave velocity profile can be inferred with resolutions on the order of 100 m and penetration depths up to several tens of kilometers. To conclude we remark that a seismic station co-located with a RLG has been installed in the underground laboratories of INFN under the Gran Sasso. The Gingerino station is now a good companion of the Wettzel observatory station. For the first time a tele seismic rotational signal has been recorded in an underground environment. The source backazimuth inferred from the directional analysis is in excellent agreement with the theoretical one, suggesting  that with a RLG and a seismometer the direction of the incoming wave-field may be estimated accurately. Corresponding to high ZLCC time intervals, we obtained estimates of phase velocities which, though being limited by the low SNR, are consistent with what expected for Love waves propagating in the PREM Earth's model \cite{DZIEWONSKI1981297} i.e. in the range of 3800 $ms^{-1}$ (at T = 10 s) to 4500 $ms^{-1}$ (at T = 50 s).
\bibliographystyle{apa}
\bibliography{bib3}


\end{multicols}

\end{document}